\documentclass[journal,letterpaper]{IEEEtran}

\usepackage{cite}
\usepackage{graphicx}
\usepackage{subfigure}
\usepackage{amssymb,amsmath} 
\usepackage{multirow}

\begin{document}

\title{Validation of proton ionization cross section generators for Monte Carlo particle transport}

\author{Matej Bati\v{c}, Maria Grazia Pia, Paolo Saracco % <-this % stops a space
\thanks{Manuscript received May 18, 2011.}% <-this % stops a space
\thanks{M. G. Pia and P. Saracco are  with
	INFN Sezione di Genova, Via Dodecaneso 33, I-16146 Genova, Italy 
	(phone: +39 010 3536328, fax: +39 010 313358,
	MariaGrazia.Pia@ge.infn.it, saracco@ge.infn.it).}
\thanks{M. Bati\v{c} is with INFN Sezione di Genova, Genova, Italy 
             (e-mail: Matej.Batic@ge.infn.it); he is on leave from 
             Jozef Stefan Institute, 1000 Ljubljana, Slovenia.}
}

\maketitle

\begin{abstract}
Three software systems, ERCS08, ISICS 2011 and \v{S}mit's code, that implement
theoretical calculations of inner shell ionization cross sections by proton
impact, are validated with respect to experimental data.
The accuracy of the cross sections they generate is quantitatively estimated
and inter-compared through statistical methods.
Updates and extensions of a cross section data library relevant to PIXE
simulation with Geant4 are discussed.

\end{abstract}

% ------------------------------------------------------------------------------
\section{Introduction}
\label{sec_intro}
\PARstart{T}{he}
calculation of inner shell ionization cross sections by proton and ion impact is 
an important component of the simulation of PIXE (Particle Induced X-ray
Emission) and the analysis of experimental PIXE spectra.
The ECPSSR (Energy-loss Coulomb Perturbed Stationary State Relativistic)
\cite{ecpssr} theory with its variants is regarded as the standard approach for
cross section calculations in the domain of PIXE applications, which typically
concern the energy range up to a few tens MeV and the whole range of
elements in the periodic system.
It provides inner shell
ionization cross sections for PIXE analysis codes such as 
GeoPIXE \cite{geopixe},
GUPIX \cite{gupix1,gupix2,gupix3},
PIXAN \cite{pixan}, 
PIXEF \cite{pixef},
PIXYKLM \cite{pixeklm}, 
Sapix \cite{sapix}, 
and TTPIXAN \cite{ttpixan},
%WinAxil \cite{winaxil} 
%and Wits-HEX \cite{witshex}, 
and for specialized PIXE simulation codes \cite{loh,viba,izarra}.

Several cross section models for the computation of inner shell ionization by
proton and $\alpha$ particle impact are available in a package for PIXE
simulation \cite{tns_pixe}  released in Geant4 \cite{g4nim,g4tns}
9.4 version; they include models based on the plane wave Born approximation
(PWBA) \cite{pwba}, the ECPSSR model in a number of variants and a collection of
empirical models, deriving from fits to experimental data.
The PWBA and ECPSSR cross section models (with variants) exploit tabulations
produced by the ISICS \cite{isics} code for K, L and M shells, which have been
assembled in a data library \cite{pixe_datalib} publicly distributed by RSICC
(Radiation Safety Information Computational Center at the Oak Ridge National
Laboratory).
%The validation of the proton cross sections included in the Geant4 PIXE data library 
%with respect to experimental data is documented in  \cite{tns_pixe}.

A new version of ISICS and an entirely new code for the calculation of ECPSSR
cross sections (with variants), ERCS08 \cite{horvat}, have become available since
the publication of the previously cited paper on Geant4 PIXE simulation.
This paper evaluates the cross sections deriving from these evolutions, along
with those calculated by \v{S}mit's code \cite{smit} (identified in the following as
KIO-LIO), which is consistent with pristine ECPSSR formulation.

The K and L shell proton ionization cross sections produced by
these three theoretical generators are compared with reference collections of
experimental data to assess their validity, in compliance with the 
IEEE Standard for Software Verification and Validation \cite{ieee_vv}.
The results of this validation process document quantitatively the relative
merits of the three codes, evaluate the impact of the newly available
calculations on Geant4 accuracy and identify the state-of-the-art of theoretical
cross sections for PIXE simulation with Geant4.

% ------------------------------------------------------------------------------
\section{Theoretical overview}

In the PWBA approach \cite{pwba}, the
first-order Born approximation is used in scattering theory to describe the
interaction between an incident charged particle and an atomic target.
This treatment is justified when the atomic number of the projectile is much
smaller than the atomic number of the target, and the velocity of the incident particle is
much larger than the velocity of the target-atom electron velocities.

The PWBA cross section in the center of mass system for the ionization of a
given shell is given by
\begin{equation}
\sigma_{\text{PWBA}} = \sigma_0 \theta^{-1}  F \left( \frac{\eta}{\theta},\theta \right)
\label{eq_pwba}
\end{equation} 
where:
\begin{equation}
\sigma_{0} = 8 \pi a_{0}^{2} \left( \frac{Z_{1}^{2}}{Z_{2}^4} \right)
\end{equation}
$a_0$ is the Bohr radius, $Z_1$ is the projectile atomic number, $Z_{2}$ is the
effective atomic number of the target atom, $F$ is the reduced universal function,
and the reduced atomic electron binding energy and reduced projectile
energy are given by
\begin{equation}
\theta = 2 n^2 \frac{U_2}{Z_2^2}
\end{equation}
\begin{equation}
\eta = 2 \frac{E_1}{M_1 Z_2^2}
\end{equation}
with E, M and U
%$E_1$, $M_1$ and $U_2$ 
representing the energy, mass and atomic binding energy
of the projectile and the target, respectively identified by the indices 1 and 2.
%In the above formulae the indices 1 and 2 refer to the projectile
%and the target respectively.
The analytical formulation of the reduced universal function $F$ can be
found in \cite{isics}.

The ECPSSR theory \cite{ecpssr} was proposed to address the shortcomings of the
PWBA approach in the energy range relevant to PIXE experimental practice
(approximately up to a few tens of MeV); it accounts for the energy loss and
Coulomb deflection of the projectile, the perturbed stationary state
and relativistic nature of the target's inner shell.
The ECPSSR cross section for a given shell is expressed in terms of the PWBA
one:
\begin{equation}
\sigma_{\text{ECPSSR}} = C_B^E(dq_{0}^{B}\zeta)\sigma_{\text{PWBA}}\left(\frac{m_R\left(\frac{\xi}{\zeta}\right)\eta}{(\zeta\theta)^2},\zeta\theta\right)
\end{equation}
where $C_E^B$ is the Coulomb deflection correction, $\zeta$ is the correction
factor for binding energy and polarization effects, $m_R$ is the relativistic
correction, $q_0$ is the minimum momentum transfer and \begin{equation}
\xi = v_1 \frac{Z_2}{U_2}
\end{equation}
$v_1$ being the projectile velocity.

Further refinements and modifications of the ECPSSR theory have been proposed:
they involve using relativistic Dirac-Hartree-Slater wave functions in computing
the form factors of the theory \cite{lapicki2005} (identified in the following
as ECPSSR-HS),
%pertinent to K shell calculations), 
account for changes in binding energy of electron states due to the presence of
the positive ion projectile in the atom (united atom correction, identified as
ECPSSR-UA), and extend the applicability of the theory to relativistic light
projectiles for K shell calculations \cite{lapicki2008} (identified as
ECPSSR-HE).

% ------------------------------------------------------------------------------
\section{Cross section generators}

The evaluation concerns three publicly accessible generators of
proton ionization cross sections.
Two of them, ISICS and ERCS08, are distributed through the Computer Physics
Communications (CPC) Program Library; the third is available directly from its
author, as specified in \cite{smit}.

% ------------------------------------------------------------------------------
\subsection{ISICS}

The ISICS code \cite{isics,isics2006,isics2008}  calculates K, L and M shell ionization cross
sections by proton and $\alpha$ particle impact according to the PWBA
approximation and the ECPSSR theory in multiple variants.

The first version of ISICS \cite{isics} provided the options of PWBA and plain
ECPSSR calculations; later versions have added capabilites for Hartree-Slater
calculations \cite{isics2006}, united atom option \cite{isics_ua} and treatment
of relativistic light projectiles \cite{isics2008}.
The latest version distributed through the CPC Program Library at the time of
writing this paper is ISICS 2011 \cite{isics2011}; it implements a few changes
with respect to the previous 2008 version, that contribute to the numerical
robustness of the code.

ISICS versions up to 2008 used Bearden and Burr's compilation \cite{bearden} of
atomic binding energies; ISICS 2011 version provides also the option
of using Williams' compilation \cite{xbook,crc90}, although the default
configuration still uses Bearden and Burr's values.

The theoretical cross sections included in the PIXE data library associated with
Geant4 9.4 have been produced with ISICS 2008.

% ------------------------------------------------------------------------------
\subsection{ERCS08}
\label{sec_ercs08}

ERCS08 \cite{horvat} is a FORTRAN program for computing electron removal cross
section by protons and heavier ions.
It encompasses the calculation of direct ionization and electron capture cross
sections; only the former is considered in this paper.

The calculations are based on the ECPSSR theory; the program allows the
configuration of individual components of the theory (E, C, PSS, R), as well as
other options: R-left (as described in \cite{lapicki2008}), united atom and hSR
(as described in \cite{lapicki2005}).

The program provides a default configuration for each shell type (K,
L or M), which can be overridden by the user.
The default configuration for the K shell activates the calculation of the form
factor involved in the ECPSSR theory using exact limits of integration and the
correction for relativistic Hartree-Slater wave functions for the K shell (the hSR option).
The united atom approach is taken into account in the default configuration for
all shells.

A few test cases documented in \cite{horvat} show that ERCS08 calculates ECPSSR
cross sections compatible with ISICS 2006 version to one unit at the fourth
significant figure, when it is configured to reproduce the same plain ECPSSR
settings as ISICS 2006.
Although \cite{horvat} states that differences with respect to ISICS may be
larger when ERCS08 is run using its own default parameters and input data, the
authors of this paper could not find evidence of the
experimental validation of the default configuration of this code in the literature.

The code includes a set of atomic binding energies; according to \cite{horvat},
they derive from experimental values in the literature when available, otherwise
from theoretical values or, when neither was deemed available, from a first
order linear regression analysis applied to available data for elements with
atomic number greater than 79 documented in \cite{horvat}.
Some of the binding energies distributed with ERCS08 could be identified as
taken from Bearden and Burr's compilation \cite{bearden}, the theoretical
values of Table IV in \cite{deslattes} and the 1978 edition of the Table of
Isotopes \cite{toi1978}; however, the source of other values remains
unidentified.
Although theoretical binding energies are available for any elements with atomic
number up to 100 in EADL (Evaluated Atomic Data Library) \cite{eadl}, they do
not appear to have been used in the program.

% ------------------------------------------------------------------------------
\subsection{KIO-LIO}

A software system to calculate ECPSSR cross sections has been developed in
Pascal by \v{Z}. \v{S}mit \cite{smit} and is available on request directly from
its author; it can be executed on Windows systems.
Since it does not appear identified in the literature by a specific name, for
convenience it is referred to in the following as KIO-LIO, as these are the
names of the executable files provided by the author respectively for K and L
shell computations.

The code implements the method described in \cite{smit} for the K shell and
$L_1$, $L_2$ subshells; the cross section for $L_3$ subshell ionization is
calculated by the same function as that for the $L_2$ subshell, with appropriate
binding energy value and further multiplied by a factor two.

The code includes three options of atomic electron binding energies,
corresponding to the compilations by Bearden and Burr \cite{bearden}, the 1978
edition of the Table of Isotopes \cite{toi1978} and Sevier \cite{sevier1979};
nevertheless, they cannot be selected through the user interface, therefore the
distributed executable files limit the production to the default configuration
with Sevier's binding energies.

%The distributed executable files were compiled with Microsoft Pascal version 3.3.

% ------------------------------------------------------------------------------
\subsection{Comparative features of calculated cross sections}

\begin{figure}
\centerline{\includegraphics[angle=0,width=8.5cm]{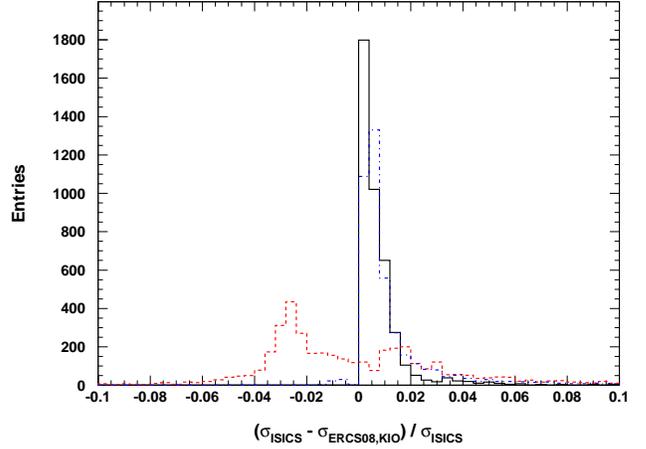}}
\caption{K shell ionization: relative difference of cross sections calculated by
default ERCS08 configuration (solid black histogram) and KIO (dashed red
histogram) with respect to ISICS ECPSSR-HS-UA configuration; relative difference
of cross sections calculated by default KIO configuration with respect to ISICS
plain ECPSSR configuration (dot-dashed blue histogram).}
\label{fig_compak}
\end{figure}

\begin{figure}
\centerline{\includegraphics[angle=0,width=8.5cm]{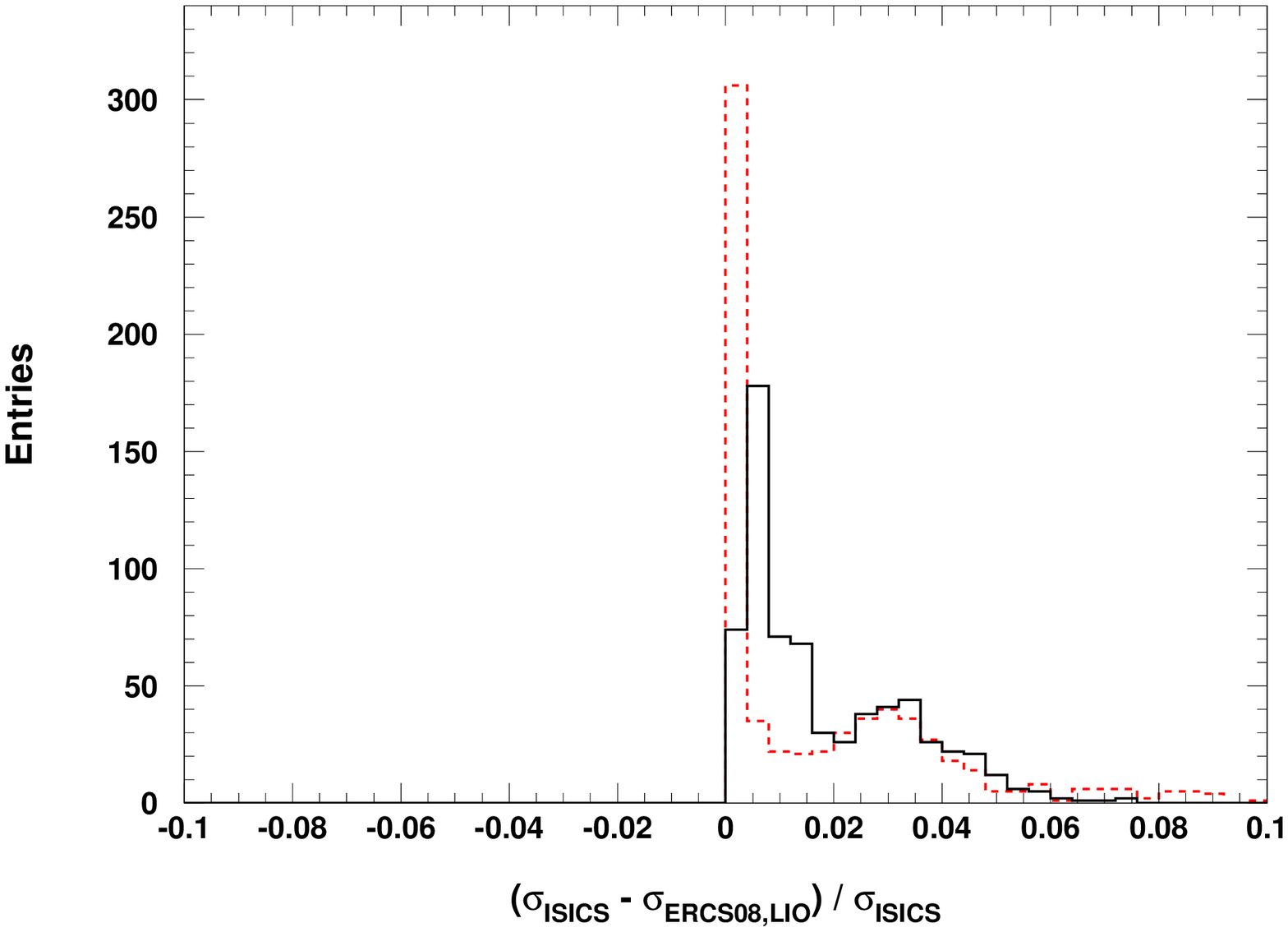}}
\caption{$L_1$ subshell ionization: relative difference of cross sections
calculated by default ERCS08 configuration (solid black histogram) and LIO
(dashed red histogram) with respect to ISICS ECPSSR-UA configuration.}
\label{fig_compal1}
\end{figure}

\begin{figure}
\centerline{\includegraphics[angle=0,width=8.5cm]{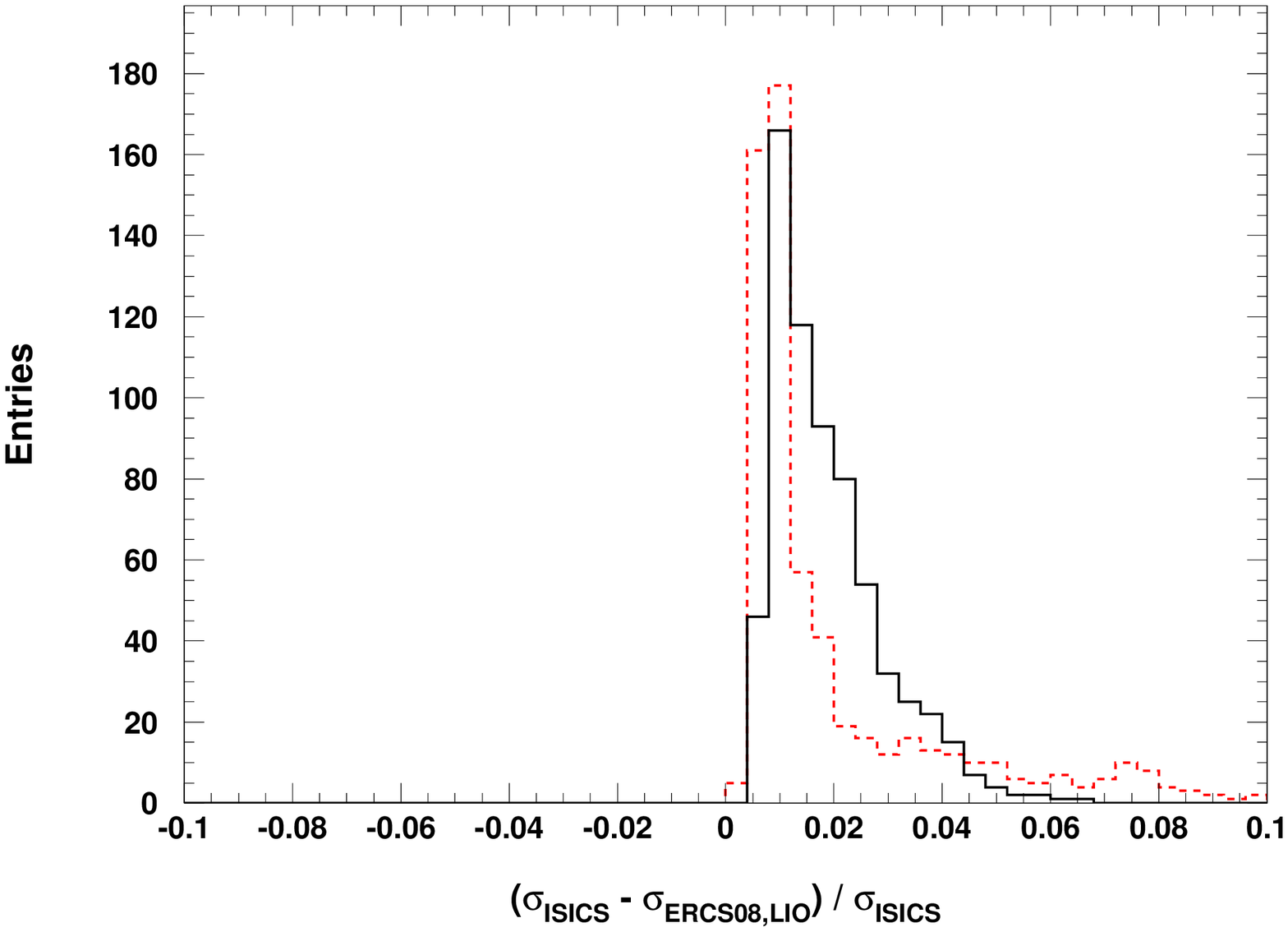}}
\caption{$L_2$ subshell ionization: relative difference of cross sections
calculated by default ERCS08 configuration (solid black histogram) and LIO
(dashed red histogram) with respect to ISICS ECPSSR-UA configuration.}
\label{fig_compal2}
\end{figure}

\begin{figure}
\centerline{\includegraphics[angle=0,width=8.5cm]{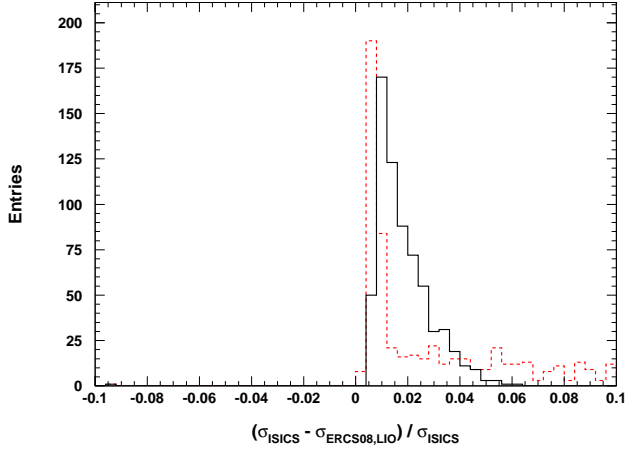}}
\caption{$L_3$ subshell ionization: relative difference of cross sections
calculated by default ERCS08 configuration (solid black histogram) and LIO
(dashed red histogram) with respect to ISICS ECPSSR-UA configuration.}
\label{fig_compal3}
\end{figure}

All of the three generators implement calculations based on the ECPSSR theory;
nevertheless, as it can be observed in Fig. \ref{fig_compak}-\ref{fig_compal3}, 
the cross sections they calculate exhibit some differences.
The figures adopt consistent color, line and marker style throughout the paper
to facilitate the identification of the behavior of each generator.

The histograms in Fig. \ref{fig_compak} concern cross sections generated by
ERCS08 and KIO in their default configurations for K shell, and by ISICS configured according
to the same nominal options as the two other codes: to calculate ECPSSR-HS-UA
cross sections (nominally equivalent to ERCS08 default) and plain ECPSSR ones
(nominally equivalent to KIO).
The cross sections are calculated at the same energies as the
experimental data collected in \cite{paul_sacher}, which, apart from a few
measurements at 160 MeV, span the energy range up to approximately 50 MeV.

Fig. \ref{fig_compal1}-\ref{fig_compal3} concerns L shell cross sections
calculated at the same energies (up to 4 MeV) as the experimental data in
\cite{orlic_exp,sokhi}: by ERCS08 and LIO in their default configurations for L
shell, and by ISICS configured to calculate ECPSSR-UA cross sections (nominally
equivalent to ERCS08 default for L subshells).
The histograms of ISICS plain ECPSSR cross sections, corresponding to LIO's
default configuration, are not superimposed to the other plots for better
clarity, since they look very close to those produced by the ECPSSR-UA option.

For nominally equivalent theoretical approaches, discrepant results may arise
from different mathematical methods adopted in the calculations, the use of
different atomic parameters (e.g. binding energies), different software
algorithms or numerical precision enforced in the code, or other computational 
details.
Comparisons with experimental data are required to quantify how the differences
of the various cross section generators affect the accuracy of the results they
produce.

%Default/optimal run configuration

% ------------------------------------------------------------------------------
\section{Validation method}

The validation process involves the comparison of cross sections calculated by
the three generators with experimental data.

% ------------------------------------------------------------------------------
\subsection{Experimental references}

The same sources of experimental data used in the validation of the Geant4 PIXE
data library documented in \cite{tns_pixe} were exploited in the validation
process of the three cross section generators considered in this paper.
They are established compilations of experimental data in the field; moreover,
the use of the same experimental references allows a comparative evaluation of
the accuracy of the newly available generators and the currently distributed
Geant4 PIXE data library, thus facilitating the identification of possible
improvements to it.

The reference experimental data for K shell ionization were extracted from the
compilation in \cite{paul_sacher}, which includes ionization cross sections
along with X-ray and Auger electron production cross sections.
The production cross sections were rescaled into ionization ones using the
fluorescence yields compiled in \cite{krause}; uncertainties were propagated accordingly.

The reference data for L shell cross sections validation were assembled from 
two complementary collections \cite{sokhi, orlic_exp}.
%The same method was applied as described for the validation of K shell cross
%sections.
The experimental compilations report only total L shell cross section data for 
a few lighter elements; sub-shell data are listed only for elements
with atomic number larger than 44.
The experimental data often exhibit large discrepancies, and systematic effects
are likely to be present for some elements, for which measurements deriving
from different sources appear to be discordant.

% ------------------------------------------------------------------------------
\subsection{Theoretical cross section production}

ISICS cross sections were calculated using ISICSoo, a refactored version of the
original ISICS Windows code, which has been especially tailored for large
scale productions of data libraries on Linux platforms.
ISICSoo provides the same physics functionality as ISICS 2011; its detailed
features and verification are documented in \cite{isics_linux}.
The source code is intended to be available through the
Computer Physics Communications Program Library.

%The experimental validation of cross sections generated by this new version of
%ISICS produces consistent results with those reported in \cite{tns_pixe}, when
%the two code versions are run in the same configuration.

\begin{figure} 
\centerline{\includegraphics[angle=0,width=8.5cm]{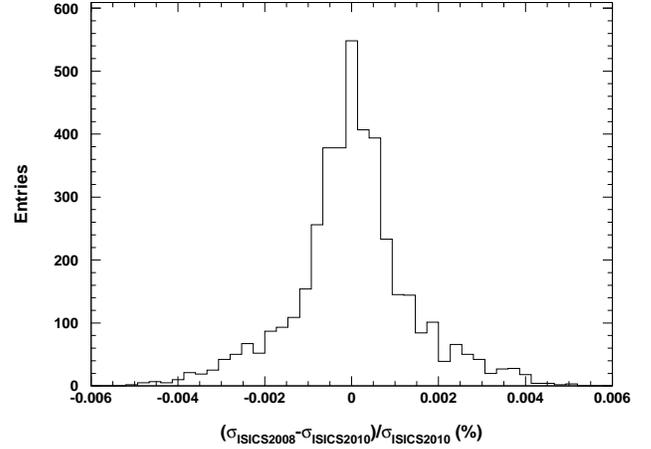}}
\caption{Relative difference of K shell ionization cross sections calculated by 
ISICS 2008 and 2011 version with the plain ECPSSR option; the latter were produced by the refactored version 
of the code identified as ISICSoo. The cross sections were computed in both cases
at incident proton energies and for target elements corresponding to the
experimental data in \cite{paul_sacher}.}
\label{fig_isicsversion}
\end{figure}

The cross sections generated by ISICSoo
appear equivalent to those computed by ISICS 2008; the relative
differences, as shown for example in Fig. \ref{fig_isicsversion} for K shell
ionization cross sections corresponding to plain ECPSSR configuration, are of
the order of $10^{-5}$.

ERCS08 was compiled and run on a Linux system to produce the cross sections
subject to validation. 
The production on a Linux platform required a small modification to the code
released through the CPC Program library, limited to the user interface for run
control. 
Input files were generated for each of the test cases through the Windows GUI to
ensure consistency with the default settings provided by the system; they
activated only the calculation of direct ionization.
The ERCS08 production system in a Linux environment was verified to generate
identical results to the original code run through the Windows GUI, 
consistent with those documented in \cite{horvat}.

The production of ISICS and ERCS08 cross sections was performed on an Intel
Core2 Quad Q9300 CPU computer equipped with Scientific Linux 5 operating 
system and gcc 4.1.2 compiler.

The executable provided by the author was used for the production of KIO-LIO
cross sections on a Microsoft Windows XP platform.

For the purpose of validation the theoretical cross sections were calculated at
the same energy as the experimental data to avoid any perturbation to the
intrinsic accuracy of the theoretical generators due to interpolation
algorithms.
This test configuration differs from the one reported in \cite{tns_pixe}, where
the cross sections subject to test were interpolated from tabulations produced
according to the pre-defined energy grid adopted for the data library associated
with the Geant4 toolkit.
The different configuration is justified by the different purpose of the test
process: validating the Geant4 cross section simulation models, that include the
interpolation of a data library, in the previous paper, and evaluating the intrinsic
accuracy of three theoretical generators in this one.

% ------------------------------------------------------------------------------
\subsection{Data analysis}
\label{sec_method}

As remarked in the previous sections, the differences among the values
calculated by the three generators are small (approximately a few percent);
their effect on the accuracy of the cross sections can be appreciated only
through statistical methods.

The analysis for the validation of theoretical cross section calculations follows
the procedure reported in \cite{tns_pixe}.
It is articulated over two stages: first a series of $\chi^2$ tests \cite{bock},
comparing theoretical and experimental cross sections for each element and shell
or subshell, followed by categorical analysis to identify differences across the
results of the $\chi^2$ test associated with the three generators.
%As described in \cite{tns_pixe}, 
Some experimental data
exhibiting large discrepancies with respect to other data at the same or
neighboring energies, suggesting the presence of systematic effects, were
excluded from the computation of the $\chi^2$ statistic.
The Statistical Toolkit \cite{gof1,gof2} was used for goodness-of-fit tests.

The null hypothesis in the goodness-of-fit testing process is defined as the
compatibility that the theoretical and experimental cross section distributions
subject to comparison derive from the same parent distribution.
Unless differently specified, a 0.05 significance level is set to define the
critical region of rejection of the null hypothesis.

%some refinements in the verification of the consistency of the
%data selection identified a few additional outliers, nevertheless the difference
%between the experimental sample in this paper and in \cite{tns_pixe} amounts to
%less than 0.2\% of the total sample size.

Contingency tables are built from the outcome of the $\chi^2$ test to
determine the equivalent behavior of the generators.
The input to contingency tables derives from the results of the $\chi^2$ test
for each element and shell or subshell: they are classified respectively as
``pass'' or ``fail'' according to whether the corresponding p-value is
consistent with the defined significance level.
The null hypothesis consists in assuming, for each contingency table, the
equivalence of the generators it compares at reproducing experimental 
measurements.
The contingency tables are analyzed with Fisher's exact test \cite{fisher}, with
the $\chi^2$ test applying Yates' continuity correction \cite{yates}, and with
Pearson's $\chi^2$ test \cite{pearson} when the number of entries in the cells
justifies its applicability.
A significance level of 0.05 is set to determine the rejection of the null
hypothesis, unless specified differently.

% ------------------------------------------------------------------------------
\section{Results}

\begin{figure} [t]
\centerline{\includegraphics[angle=0,width=8.5cm]{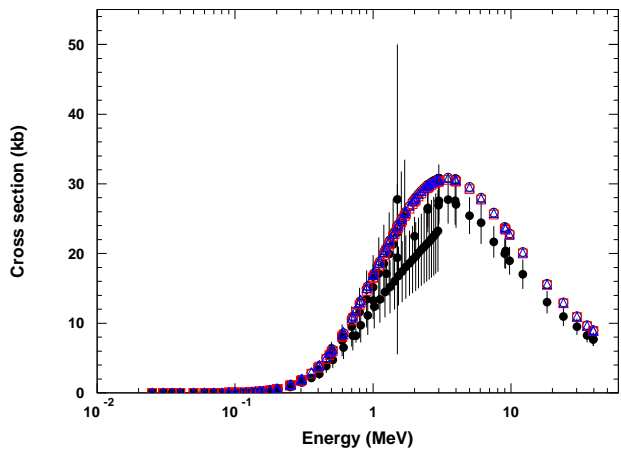}}
\caption{K shell ionization cross sections by proton impact on aluminium
calculated by theoretical generators and experimental data from
\cite{paul_sacher} (black filled circles): ISICS 2011 in ECPSSR-HS-UA
configuration (empty circles), ERCS08 (empty squares) and KIO (empty triangles), both
in default configuration.}
\label{fig_crossk_13}
\end{figure}

\begin{figure} [t]
\centerline{\includegraphics[angle=0,width=8.5cm]{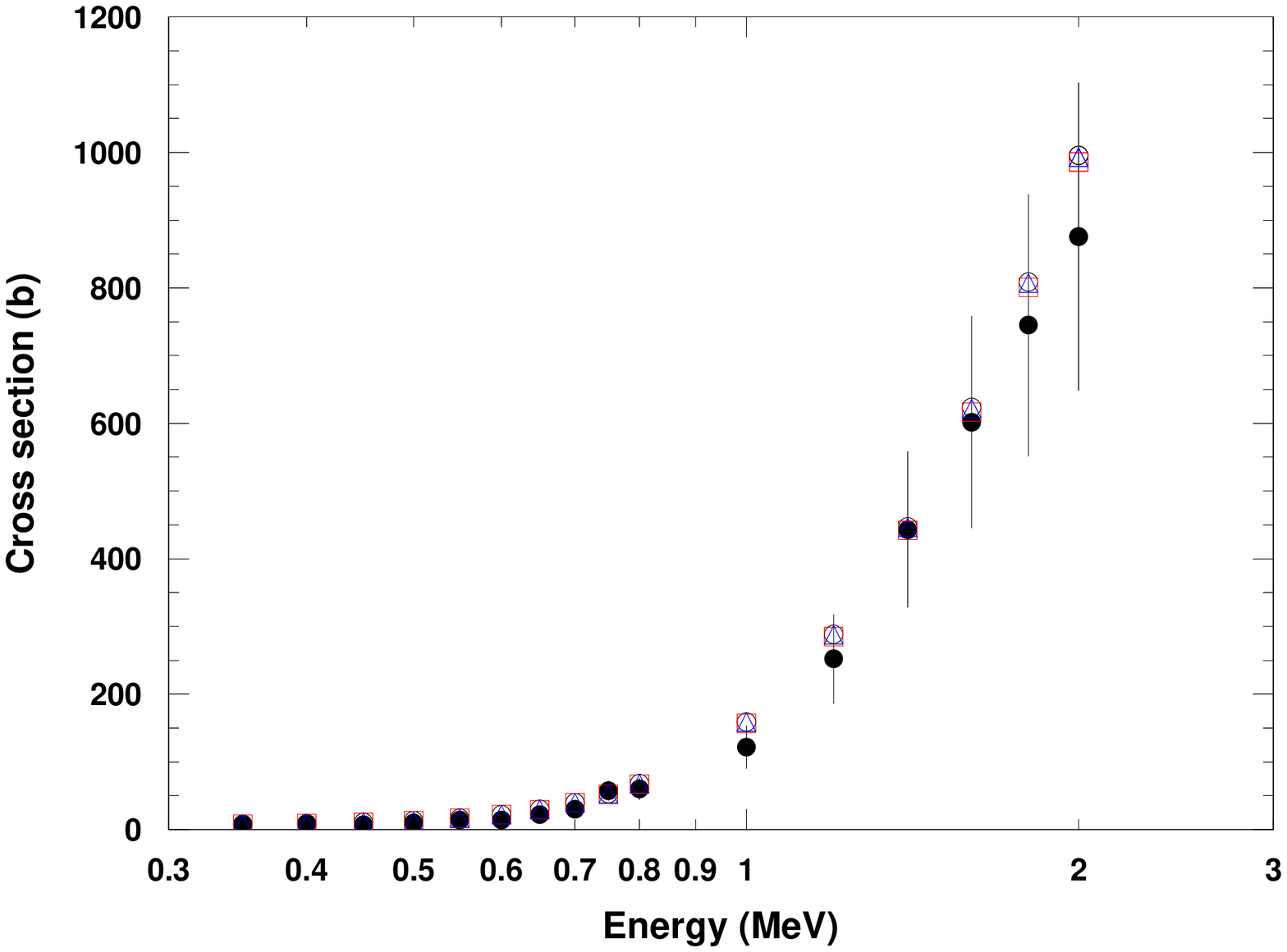}}
\caption{$L_1$ subshell ionization cross sections by proton impact on tellurium
calculated by theoretical generators and experimental data from
\cite{orlic_exp,sokhi} (black filled circles): ISICS 2011 in ECPSSR-UA
configuration (empty circles), ERCS08 (empty squares) and LIO (empty triangles), both
in default configuration.}
\label{fig_crossl1_52}
\end{figure}

\begin{figure} [th!]
\centerline{\includegraphics[angle=0,width=8.5cm]{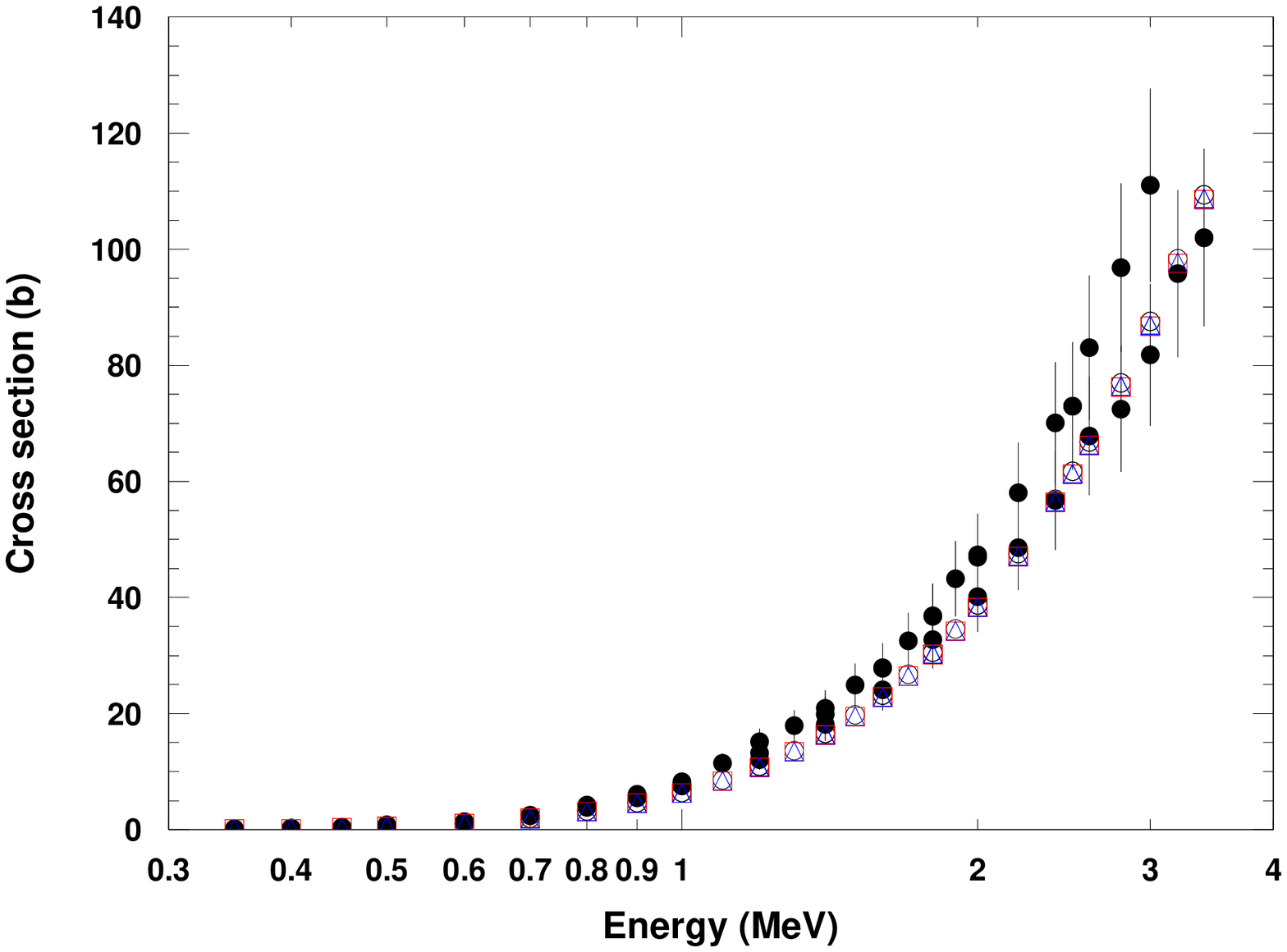}}
\caption{$L_2$ subshell ionization cross sections by proton impact on tungsten
calculated by  theoretical generators and experimental data from
\cite{orlic_exp,sokhi} (black filled circles): ISICS 2011 in ECPSSR-UA
configuration (empty circles), ERCS08 (empty squares) and LIO (empty triangles), both
in default configuration.}
\label{fig_crossl2_74}
\end{figure}

\begin{figure} [th!]
\centerline{\includegraphics[angle=0,width=8.5cm]{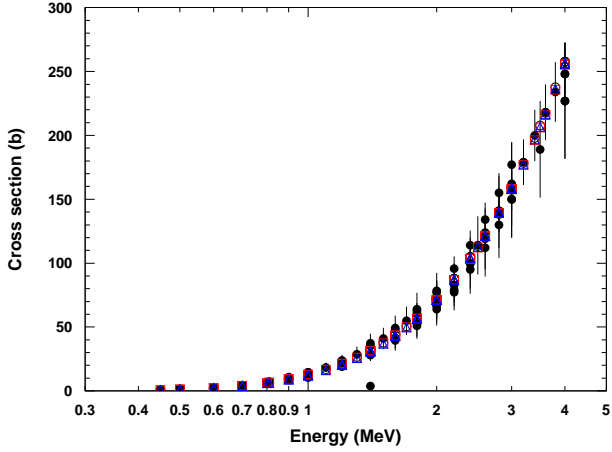}}
\caption{$L_3$ subshell ionization cross sections by proton impact on gold
calculated by  theoretical generators and experimental data from
\cite{orlic_exp,sokhi} (black filled circles): ISICS 2011 in ECPSSR-UA
configuration (empty circles), ERCS08 (empty squares) and LIO (empty triangles), both
in default configuration.}
\label{fig_crossl3_79}
\end{figure}

A selection of experimental data and cross sections calculated by the three
generators is displayed in Fig. \ref{fig_crossk_13}-\ref{fig_crossl3_79}; the
theoretical values correspond to the generators' configurations analyzed
in the following sections.

\begin{figure}
\centerline{\includegraphics[angle=0,width=8.5cm]{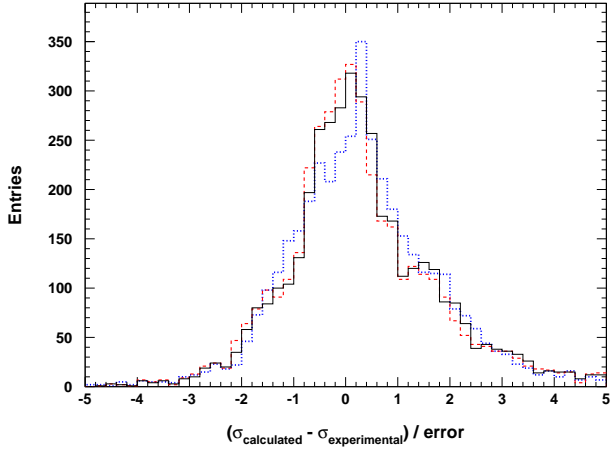}}
\caption{Difference between K shell cross sections calculated by the three
generators and experimental data, divided by the related experimental
uncertainties: ISICS 2011 in ECPSSR-HS-UA configuration (solid black histogram),
ERCS08 (red dashed histogram) and KIO (dotted blue histogram) in default
configuration.}
\label{fig_nsigmak}
\end{figure}

\begin{figure}
\centerline{\includegraphics[angle=0,width=8.5cm]{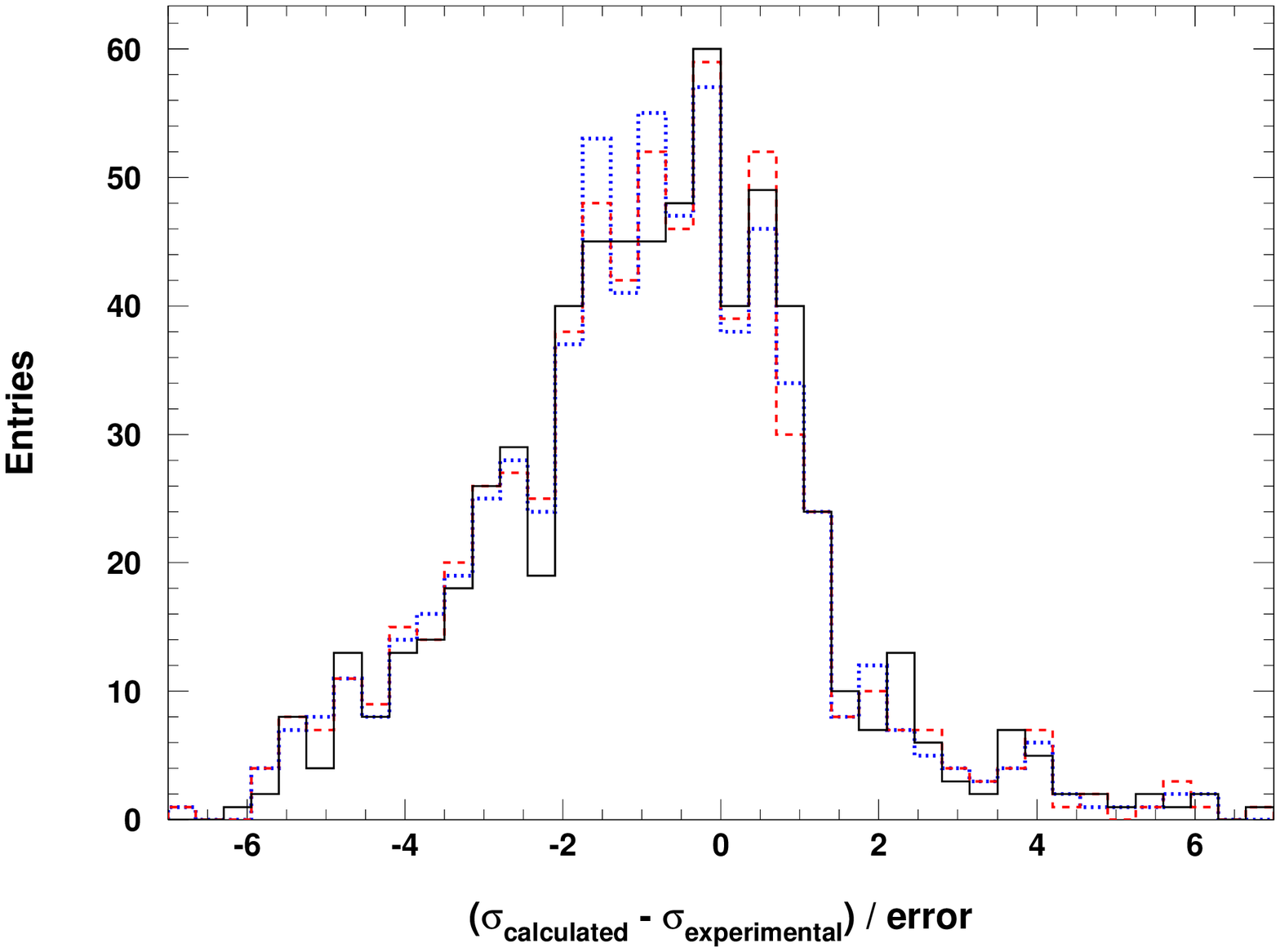}}
\caption{Difference between $L_1$ subshell cross sections calculated by the
three generators and experimental data, divided by the related experimental
uncertainties: ISICS 2011 in ECPSSR-UA configuration (solid black histogram),
ERCS08 (red dashed histogram) and LIO (dotted blue histogram) in default
configuration.}
\label{fig_nsigmal1}
\end{figure}

\begin{figure}
\centerline{\includegraphics[angle=0,width=8.5cm]{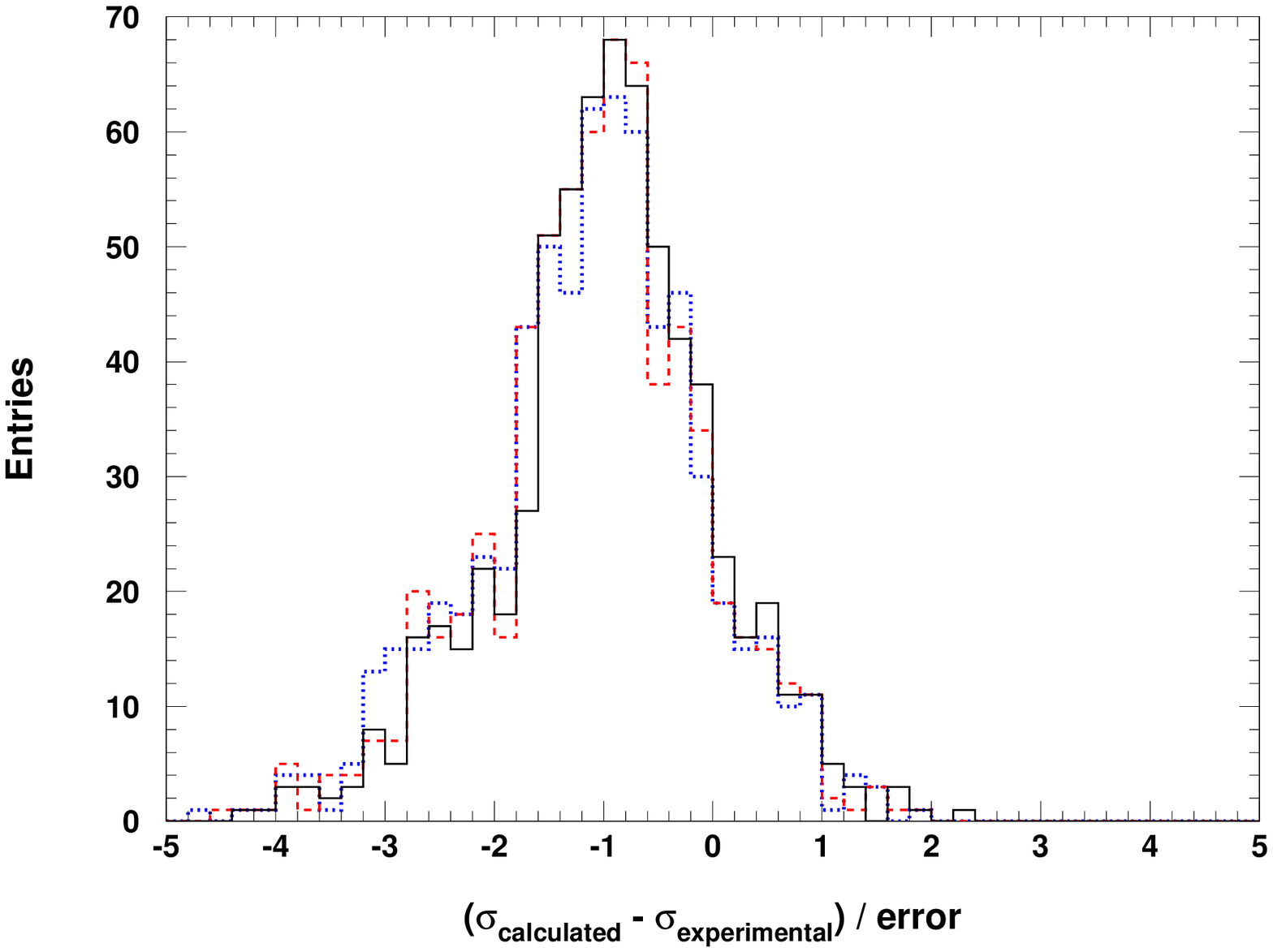}}
\caption{Difference between $L_2$ subshell cross sections calculated by the
three generators and experimental data, divided by the related experimental
uncertainties: ISICS 2011 in ECPSSR-UA configuration (solid black histogram),
ERCS08 (red dashed histogram) and LIO (dotted blue histogram) in default
configuration.}
\label{fig_nsigmal2}
\end{figure}

\begin{figure}
\centerline{\includegraphics[angle=0,width=8.5cm]{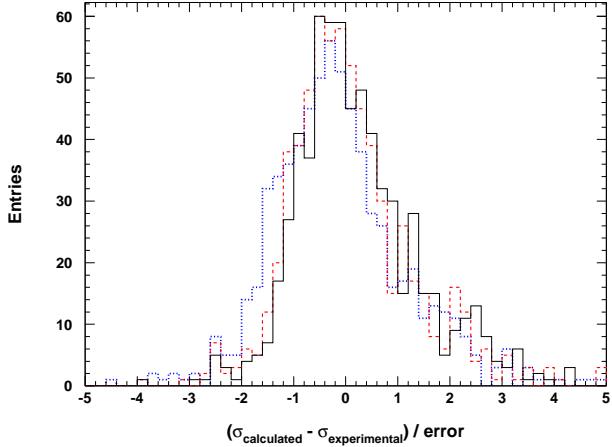}}
\caption{Difference between $L_3$ subshell cross sections calculated by the
three generators and experimental data, divided by the related experimental
uncertainties: ISICS 2011 in ECPSSR-UA configuration (solid black histogram),
ERCS08 (red dashed histogram) and LIO (dotted blue histogram) in default
configuration.}
\label{fig_nsigmal3}
\end{figure}

A comprehensive overview of how theoretical cross sections compare to
measurements is shown in Fig. \ref{fig_nsigmak}-\ref{fig_nsigmal3}: the plots
display the
difference between the cross sections calculated by the three generators and
experimental data, divided by the corresponding experimental uncertainties.
In other words, they show by how many standard deviations experimental 
values differ from theoretical ones.
These plots include all the experimental data of \cite{paul_sacher,orlic_exp,sokhi},
without excluding any outliers.
Theoretical generators appear to systematically underestimate $L_1$ and
$L_2$ experimental cross sections.

The following sections report the statistical analysis to evaluate in detail the
compatibility of the theoretical cross sections with experimental measurements.
It is worthwhile to note that the incompatibility with experiment of all generators 
in some test cases hints to either an intrinsic deficiency of the underlying theory or to
systematic effects in the experimental data.

% ------------------------------------------------------------------------------
\subsection{K shell}
\label{sec_k}

The p-values of the $\chi^2$ test comparing calculated and experimental K
cross sections are reported for each tested element in Table \ref{tab_pvaluek}.
The results are listed for ERCS08 and KIO in their default configuration, for
ISICS 2011 in various options, two of which, ECPSSR-HS-UA and plain ECPSSR, are 
similar to the default configurations of the other codes.

The results are summarized in Table \ref{tab_effk}, where the test cases for
which the null hypothesis of compatibility of calculated and experimental
distributions is rejected at 0.05 significance level are counted as ``fail'',
otherwise as ``pass''.
The efficiency of each theoretical generator's configuration is defined as the ratio
between the ``passed'' and total test cases.

It is evident from the results in Table \ref{tab_effk} that ISICS 2011 ECPSSR-HS,
ECPSSR-HS-UA and ERCS08 default configuration are the most efficient at
reproducing experimental cross sections, while KIO and the other ISICS options
produce slightly less accurate results.
Nevertheless, as shown in the contingency tables in Table \ref{tab_contk}, the
differences in efficiency are not statistically significant at 95\% confidence
level.

Although the statistical analysis over the complete data samples provides an
overall picture of the accuracy of the three cross section generators, the
detailed results of the $\chi^2$ test in Table \ref{tab_pvaluek} can be useful
to optimize the choice in some specialized use cases: for instance, ERCS08 cross
sections for silicon are compatible with experimental data \cite{paul_sacher}, while the two other
generators fail at reproducing compatible cross sections at 0.05 significance
level for this element; KIO succeeds at reproducing equivalent tungsten
cross sections at 0.05 significance level, while both ISICS ECPSSR-HS-UA and ERCS08
default configurations fail.
In this respect, however, it is worthwhile to note that in the case of tungsten,
as well as for other test cases concerning heavy elements and relatively high
($\gtrsim 5-7$ MeV) energy, the Hartree-Slater correction does not improve the accuracy
of ECPSSR calculations, actually the opposite.
%, as already remarked in \cite{tns_pixe}.

\begin{table*}
\begin{center}
\caption{P-values of the $\chi^2$ test comparing calculated and experimental K
shell ionization cross sections by proton impact}
\label{tab_pvaluek}
\begin{tabular}{|l|ccccc|c|c|}
\hline
        &\multicolumn{5}{|c|}{\bf ISICS 2011}                                &{\bf ERCS08} & {\bf KIO}   \\
Z	&ECPSSR & ECPSSR-HS	& ECPSSR-UA	& ECPSSR-HE	& ECPSSR-HS-UA	&Default  &Default     \\
\hline
6	& 0.751	& 0.944	& 0.751	& 0.774	& 0.944	& 0.985	& 0.860  \\
7	& $<0.001$	& 0.079	& $<0.001$	& $<0.001$	& 0.079	& 0.885	& $<0.001$  \\
8	& 0.986	& 0.986	& 0.983	& 0.986	& 0.987	& 0.945	& 0.986  \\
9	& 0.800	& 0.733	& 0.800	& 0.784	& 0.733	& 0.943	& 0.809  \\
10	& 0.999	& 0.996	& 0.999	& 0.999	& 0.996	& 0.995	& 1.000  \\
12	& 0.039	& 0.551	& 0.012	& 0.040	& 0.344	& 0.725	& 0.057  \\
13	& 0.003	& 0.079	& 0.001	& $<0.001$	& 0.061	& 0.292	& 0.004  \\
14	& 0.014	& 0.029	& 0.014	& 0.013	& 0.029	& 0.096	& 0.016  \\
15	& 0.710	& 0.612	& 0.710	& 0.706	& 0.612	& 0.515	& 0.695  \\
16	& 0.802	& 0.719	& 0.802	& 0.801	& 0.719	& 0.632	& 0.790  \\
17	& 0.650	& 0.732	& 0.650	& 0.657	& 0.732	& 0.787	& 0.666  \\
18	& 0.403	& 0.815	& 0.599	& 0.434	& 0.945	& 0.938	& 0.448  \\
19	& 0.993	& 0.989	& 0.993	& 0.993	& 0.989	& 0.985	& 0.993  \\
20	& 0.029	& 0.117	& 0.029	& 0.020	& 0.117	& 0.178	& 0.036  \\
21	& 0.257	& 1.000	& 0.257	& 0.331	& 1.000	& 1.000	& 0.490  \\
22	& 0.126	& 0.118	& 0.123	& 0.142	& 0.128	& 0.046	& 0.154  \\
23	& 0.059	& 0.517	& 0.053	& 0.082	& 0.499	& 0.537	& 0.111  \\
24	& 0.374	& 0.848	& 0.344	& 0.448	& 0.830	& 0.832	& 0.511  \\
25	& 0.405	& 0.526	& 0.276	& 0.417	& 0.387	& 0.373	& 0.508  \\
26	& 0.987	& 0.739	& 0.947	& 0.987	& 0.572	& 0.365	& 0.988  \\
27	& 0.969	& 0.983	& 0.965	& 0.980	& 0.980	& 0.936	& 0.982  \\
28	& 0.177	& 0.491	& 0.096	& 0.158	& 0.204	& 0.314	& 0.225  \\
29	& 0.653	& 0.890	& 0.504	& 0.749	& 0.811	& 0.806	& 0.812  \\
30	& 0.997	& 0.990	& 0.994	& 0.997	& 0.984	& 0.971	& 0.999  \\
31	& 0.042	& 0.381	& 0.042	& 0.075	& 0.381	& 0.465	& 0.083  \\
32	& 0.001	& 0.138	& 0.001	& 0.004	& 0.120	& 0.208	& 0.006  \\
33	& 0.066	& 0.392	& 0.066	& 0.104	& 0.392	& 0.496	& 0.143  \\
34	& 0.409	& 0.629	& 0.411	& 0.482	& 0.639	& 0.641	& 0.464  \\
35	& 0.248	& 0.312	& 0.248	& 0.291	& 0.312	& 0.297	& 0.249  \\
36	& 0.235	& 0.522	& 0.235	& 0.320	& 0.522	& 0.538	& 0.383  \\
37	& 0.105	& 0.147	& 0.105	& 0.119	& 0.147	& 0.143	& 0.087  \\
38	& 0.019	& 0.023	& 0.019	& 0.025	& 0.023	& 0.031	& 0.049  \\
39	& 0.109	& 0.189	& 0.109	& 0.121	& 0.189	& 0.205	& 0.147  \\
40	& 0.039	& 0.067	& 0.037	& 0.056	& 0.065	& 0.080	& 0.088  \\
41	& 0.001	& 0.002	& 0.002	& 0.003	& 0.003	& 0.003	& 0.002  \\
42	& 0.998	& 0.999	& 0.998	& 1.000	& 0.999	& 0.999	& 0.999  \\
45	& $<0.001$	& $<0.001$	& $<0.001$	& $<0.001$	& $<0.001$	& $<0.001$	& $<0.001$  \\
46	& 0.621	& 0.981	& 0.573	& 0.614	& 0.936	& 0.957	& 0.851  \\
47	& 0.332	& 0.379	& 0.332	& 0.562	& 0.398	& 0.378	& 0.589  \\
48	& 0.001	& $<0.001$	& 0.001	& 0.007	& $<0.001$	& $<0.001$	& 0.004  \\
49	& $<0.001$	& $<0.001$	& $<0.001$	& $<0.001$	& $<0.001$	& $<0.001$	& $<0.001$  \\
50	& 0.946	& 0.957	& 0.946	& 0.942	& 0.957	& 0.962	& 0.980  \\
51	& 0.787	& 0.874	& 0.787	& 0.812	& 0.874	& 0.889	& 0.961  \\
52	& $<0.001$	& $<0.001$	& $<0.001$	& $<0.001$	& $<0.001$	& $<0.001$	& $<0.001$  \\
53	& 0.057	& 0.014	& 0.057	& 0.119	& 0.014	& 0.018	& 0.305  \\
55	& 0.327	& 0.468	& 0.327	& 0.247	& 0.468	& 0.468	& 0.280  \\
56	& 0.310	& 0.922	& 0.310	& 0.147	& 0.922	& 0.914	& 0.248  \\
57	& 0.056	& 0.170	& 0.056	& 0.037	& 0.170	& 0.166	& 0.052  \\
58	& 0.073	& 0.579	& 0.073	& 0.063	& 0.579	& 0.588	& 0.378  \\
59	& 0.010	& 0.009	& 0.010	& 0.014	& 0.009	& 0.009	& 0.031  \\
60	& 0.243	& 0.596	& 0.243	& 0.148	& 0.596	& 0.609	& 0.355  \\
62	& 0.218	& 0.976	& 0.218	& 0.110	& 0.976	& 0.978	& 0.514  \\
63	& $<0.001$	& $<0.001$	& 0.001	& 0.002	& $<0.001$	& $<0.001$	& 0.010  \\
64	& 0.394	& 0.324	& 0.394	& 0.082	& 0.324	& 0.340	& 0.500  \\
65	& 0.008	& 0.522	& 0.008	& 0.003	& 0.522	& 0.527	& 0.025  \\
67	& 0.020	& 0.997	& 0.020	& 0.001	& 0.998	& 0.997	& 0.011  \\
69	& 0.014	& 0.014	& 0.014	& 0.022	& 0.007	& 0.015	& 0.049  \\
70	& $<0.001$	& $<0.001$	& $<0.001$	& 0.008	& $<0.001$	& $<0.001$	& $<0.001$  \\
72	& 0.275	& 0.647	& 0.275	& 0.311	& 0.647	& 0.650	& 0.645  \\
73	& 0.029	& 0.004	& 0.029	& 0.019	& 0.004	& 0.005	& 0.018  \\
74	& 0.232	& $<0.001$	& 0.232	& 0.350	& $<0.001$	& $<0.001$	& 0.719  \\
75	& 0.075	& 0.070	& 0.075	& 0.093	& 0.070	& 0.072	& 0.185  \\
78	& 0.186	& 0.528	& 0.186	& 0.187	& 0.092	& 0.094	& 0.001  \\
79	& 0.107	& $<0.001$	& 0.107	& 0.454	& $<0.001$	& $<0.001$	& 0.667  \\
82	& 0.003	& $<0.001$	& 0.003	& 0.128	& $<0.001$	& $<0.001$	& 0.022  \\
83	& 0.001	& 0.256	& 0.001	& 0.001	& 0.256	& 0.265	& 0.052  \\
90	& 0.007	& $<0.001$	& 0.007	& 0.003	& $<0.001$	& $<0.001$	& 0.001  \\
92	& 0.136	& 0.646	& 0.136	& 0.158	& 0.646	& 0.657	& 0.524  \\
\hline
\end{tabular}
\end{center}
\end{table*}							         

\begin{table*}
\begin{center}
\caption{Summary of the $\chi^2$ test results of K shell ionization cross
sections by proton impact}
\label{tab_effk}
\begin{tabular}{|l|ccccc|c|c|}
\hline
        &\multicolumn{5}{|c|}{\bf ISICS 2011}                                &{\bf ERCS08} &{\bf KIO}   \\
	&ECPSSR & ECPSSR-HS	& ECPSSR-UA	& ECPSSR-HE	& ECPSSR-HS-UA	&Default &Default     \\
\hline
Tested elements	& 66	& 66	& 66	& 66	& 66	& 66	& 66  \\
Pass		& 44	& 51	& 44	& 46	& 51	& 51	& 47  \\
Fail		& 22	& 15	& 22	& 20	& 15	& 15	& 19  \\
Efficiency  	& 0.67$\pm$0.06	&0.77$\pm$0.05	&0.67$\pm$0.06	&0.70$\pm$0.06	&0.77$\pm$0.05	&0.77$\pm$0.05	&0.71$\pm$0.06 \\
\hline
\end{tabular}
\end{center}
\end{table*}							         

\begin{table} [bpht]
\begin{center}
\caption{Contingency table to estimate the equivalent accuracy of ECPSSR K shell
cross sections using EADL and Bearden and Burr's binding energies}
\label{tab_contk}
\begin{tabular}{|l|c|c|c|c|}
\hline
{\bf $\chi^2$ test outcome}		&{\bf ISICS 2011} &{\bf ERCS08}	\\
\hline
Pass					& 51			& 51 \\
Fail					& 15			& 15 \\
\hline
p-value Fisher test		& \multicolumn{2}{|c|}{1} \\
p-value Pearson $\chi^2$	& \multicolumn{2}{|c|}{1} \\
p-value Yates $\chi^2$		& \multicolumn{2}{|c|}{0.835} \\
\hline
{\bf $\chi^2$ test outcome}	&{\bf ISICS 2011} &{\bf KIO}	\\
\hline
Pass					& 51			& 47 \\
Fail					& 15			& 19 \\
\hline
p-value Fisher test		& \multicolumn{2}{|c|}{0.551} \\
p-value Pearson $\chi^2$	& \multicolumn{2}{|c|}{0.426} \\
p-value Yates $\chi^2$		& \multicolumn{2}{|c|}{0.550} \\
\hline
\end{tabular}
\end{center}
\end{table}

% ------------------------------------------------------------------------------
\subsection{L shell}

The p-values of the $\chi^2$ test comparing calculated and experimental $L_1$,
$L_2$ and $L_3$ subshell cross sections are reported for each tested element in
Tables \ref{tab_pvaluel1}-\ref{tab_pvaluel3}.
Whenever the data for a given element were present in both experimental
compilations \cite{orlic_exp,sokhi}, only the results associated with the
compilation exhibiting the larger p-values are listed in the tables.
The results are listed for ERCS08 and LIO in their default configuration, for
ISICS 2011 in the ECPSRR-UA and plain ECPSSR configurations, which are 
similar to the default configurations of the other codes.

The results are summarized in Table \ref{tab_effl}, where the entries for
``pass'', ``fail'' and efficiency correspond to the same definitions as
in the previous section.
ISICS 2011 ECPSSR-UA is the most efficient at reproducing experimental cross
sections, nevertheless, as shown in the contingency tables in
Table \ref{tab_contl}, the differences in efficiency are not statistically
significant at 95\% confidence level.

\begin{table} 
\begin{center}
\caption{P-values of the $\chi^2$ test comparing calculated and experimental
$L_1$ subshell ionization cross sections by proton impact}
\label{tab_pvaluel1}
\begin{tabular}{|l|cc|c|c|}
\hline
        &\multicolumn{2}{|c|}{\bf ISICS 2011} &{\bf ERCS08} & {\bf LIO}   \\
Z	&ECPSSR	&ECPSSR-UA	&Default	&Default  \\
\hline
45	& 0.030	& 0.004	& 0.021	& 0.079  \\
46	& 0.598	& 0.528	& 0.678	& 0.686  \\
47	& 0.966	& 0.951	& 0.971	& 0.979  \\
48	& 0.850	& 0.858	& 0.865	& 0.862  \\					
49	& 0.067	& 0.087	& 0.056	& 0.056  \\
50	& 0.197	& 0.122	& 0.275	& 0.344  \\
51	& 0.946	& 0.972	& 0.936	& 0.910  \\					
52	& 0.474	& 0.419	& 0.563	& 0.533  \\
59	& 0.056	& 0.101	& 0.040	& 0.022  \\
62	& 0.981	& 0.977	& 0.992	& 0.988  \\
64	& 0.263	& 0.243	& 0.236	& 0.212  \\
65	& 0.974	& 0.990	& 0.924	& 0.865  \\
66	& 0.951	& 0.953	& 0.939	& 0.938  \\					
67	& 0.126	& 0.127	& 0.085	& 0.084  \\					
70	& 0.619	& 0.619	& 0.553	& 0.552  \\					
72	& 0.333	& 0.333	& 0.418	& 0.389  \\					
73	& 0.929	& 0.929	& 0.928	& 0.911  \\
74	& $<0.001$	& $<0.001$	& $<0.001$	& $<0.001$  \\					
75	& 0.975	& 0.975	& 0.986	& 0.986  \\
76	& 0.020	& 0.020	& 0.003	& 0.003  \\
77	& $<0.001$	& $<0.001$	& $<0.001$	& $<0.001$  \\				
78	& $<0.001$	& $<0.001$	& $<0.001$	& $<0.001$  \\
79	& 0.916	& 0.916	& 0.873	& 0.877  \\
81	& $<0.001$	& $<0.001$	& $<0.001$	& $<0.001$  \\
82	& $<0.001$	& $<0.001$	& $<0.001$	& $<0.001$  \\
83	& 0.124	& 0.124	& 0.076	& 0.082  \\					
90	& 0.003	& 0.003	& 0.006	& 0.007  \\
92	& $<0.001$	& $<0.001$	& $<0.001$	& $<0.001$  \\
\hline
\end{tabular}
\end{center}
\end{table}	

\begin{table} 
\begin{center}
\caption{P-values of the $\chi^2$ test comparing calculated and experimental
$L_2$ subshell ionization cross sections by proton impact}
\label{tab_pvaluel2}
\begin{tabular}{|l|cc|c|c|}
\hline
        &\multicolumn{2}{|c|}{\bf ISICS 2011} &{\bf ERCS08} & {\bf LIO}   \\
Z	&ECPSSR	&ECPSSR-UA	&Default	&Default  \\
\hline
45      & 0.001	& 0.057	& 0.023	& $<0.001$  \\
46      & $<0.001$	& 0.008	& 0.002	& $<0.001$  \\
47      & 0.015	& 0.139	& 0.067	& 0.029  \\
48      & 0.101	& 0.121	& 0.080	& 0.084  \\
49      & 0.082	& 0.256	& 0.136	& 0.055  \\
50      & 0.092	& 0.163	& 0.132	& 0.073  \\			
51      & 1.000	& 1.000	& 1.000	& 1.000  \\
52      & 0.718	& 0.722	& 0.822	& 0.761  \\
59      & 0.350	& 0.654	& 0.581	& 0.258  \\
62      & 0.005	& 0.060	& 0.026	& 0.002  \\
64      & $<0.001$	& $<0.001$	& $<0.001$	& $<0.001$  \\
65      & $<0.001$	& $<0.001$	& $<0.001$	& $<0.001$  \\
66      & 0.836	& 0.862	& 0.821	& 0.777  \\			
67      & 0.173	& 0.312	& 0.196	& 0.093  \\			
70      & 0.129	& 0.139	& 0.065	& 0.064  \\			
72      & 0.984	& 0.991	& 0.967	& 0.962  \\			
73      & 0.054	& 0.116	& 0.065	& 0.028  \\			
74      & 0.341	& 0.453	& 0.231	& 0.185  \\
75      & 0.398	& 0.456	& 0.294	& 0.283  \\
76      & 0.762	& 0.871	& 0.677	& 0.556  \\			
77      & 0.331	& 0.433	& 0.208	& 0.166  \\
78      & 0.655	& 0.713	& 0.500	& 0.478  \\
79      & 1.000	& 1.000	& 0.998	& 0.999  \\			
81      & 0.018	& 0.018	& 0.009	& 0.010  \\
82      & 0.841	& 0.843	& 0.711	& 0.728  \\			
83      & 0.322	& 0.322	& 0.187	& 0.187  \\						
90      & 0.049	& 0.049	& 0.023	& 0.025  \\
92      & $<0.001$	& $<0.001$	& $<0.001$	& $<0.001$  \\
\hline
\end{tabular}
\end{center}
\end{table}	

\begin{table} 
\begin{center}
\caption{P-values of the $\chi^2$ test comparing calculated and experimental
$L_3$ subshell ionization cross sections by proton impact}
\label{tab_pvaluel3}
\begin{tabular}{|l|cc|c|c|}
\hline
        &\multicolumn{2}{|c|}{\bf ISICS 2011} &{\bf ERCS08} & {\bf LIO}   \\
Z	&ECPSSR	&ECPSSR-UA	&Default	& Default \\
\hline
45      & 0.235	& 0.987	& 0.948	& 0.168  \\
46      & 0.061	& 0.829	& 0.620	& 0.038  \\
47      & 0.436	& 0.969	& 0.888	& 0.347  \\			
48      & 0.909	& 0.949	& 0.919	& 0.891  \\
49      & 0.679	& 0.978	& 0.907	& 0.583  \\
50      & 0.221	& 0.137	& 0.172	& 0.204  \\			
51      & 0.999	& 0.999	& 1.000	& 0.999  \\
52      & 0.535	& 0.463	& 0.608	& 0.594  \\
59      & 0.019	& 0.399	& 0.245	& 0.010  \\
62      & 0.087	& 0.764	& 0.586	& 0.044  \\
64      & 0.783	& 0.274	& 0.581	& 0.679  \\
65      & 0.309	& 0.912	& 0.833	& 0.188  \\
66      & 0.643	& 0.502	& 0.589	& 0.592  \\			
67      & 0.102	& 0.512	& 0.402	& 0.048  \\
70      & 0.599	& 0.991	& 0.963	& 0.394  \\						
72      & 0.970	& 0.950	& 0.992	& 0.984  \\
73      & 0.525	& 0.867	& 0.652	& 0.339  \\			
74      & 0.701	& 0.928	& 0.807	& 0.549  \\			
75      & 0.960	& 0.960	& 0.992	& 0.971  \\
76      & $<0.001$	& $<0.001$	& 0.004	& $<0.001$  \\
77      & 1.000	& 0.979	& 0.999	& 1.000  \\			
78      & $<0.001$	& $<0.001$	& $<0.001$	& $<0.001$  \\
79      & 1.000	& 1.000	& 1.000	& 1.000  \\			
81      & 0.060	& 0.109	& 0.066	& 0.038  \\
82      & 1.000	& 1.000	& 1.000	& 1.000  \\			
83      & 0.460	& 0.508	& 0.560	& 0.430  \\			
90      & 0.963	& 0.994	& 0.978	& 0.907  \\			
92      & 0.085	& 0.021	& 0.080	& 0.179  \\
\hline
\end{tabular}
\end{center}
\end{table}

\begin{table*}
\begin{center}
\caption{Summary of $\chi^2$ test results of L subshell ionization cross
sections by proton impact}
\label{tab_effl}
\begin{tabular}{|ll|cc|c|c|}
\hline				         
\multicolumn{2}{|c|}{}	&\multicolumn{2}{|c|}{\bf ISICS 2011} &{\bf ERCS08} & {\bf LIO}   \\
\multicolumn{2}{|c|}{}   &ECPSSR	&ECPSSR-UA	&Default	&Default  \\
\hline
\multirow{4} {*} {$L_1$}
&Elements  	&28	& 28	& 28	& 28     \\
&Pass            	& 19	& 19	& 18	& 19     \\
&Fail            	& 9	& 9	& 10	& 9     \\
&Efficiency      	& 0.53$\pm$0.09	& 0.53$\pm$0.09	& 0.48$\pm$0.09	& 0.50$\pm$0.09   \\
\hline
\multirow{4} {*} {$L_2$}
&Elements &28	& 28	& 28	& 28     \\
&Pass            	& 19	& 22	& 20	& 18     \\
&Fail            	& 9	& 6	& 8	& 10     \\
&Efficiency      	& 0.68$\pm$0.09	& 0.79$\pm$0.08	& 0.71$\pm$0.09	& 0.64$\pm$0.09   \\
\hline
\multirow{4} {*} {$L_3$}
&Elements &28	& 28	& 28	& 28     \\
&Pass            	& 25	& 25	& 26	& 21     \\
&Fail            	& 3	& 3	& 2	& 7      \\
&Efficiency      	& 0.89$\pm$0.06	& 0.89$\pm$0.06	& 0.93$\pm$0.05	& 0.75$\pm$0.08   \\
\hline
\multirow{4} {*} {$L$}
&Elements 		& 84	& 84	& 84	& 84     \\
&Pass            	& 63	& 66	& 64	& 58     \\
&Fail            		& 21	& 18	& 20	& 26     \\
&Efficiency      	& 0.75$\pm$0.05	& 0.79$\pm$0.04	& 0.76$\pm$0.59	& 0.69$\pm$0.05   \\
\hline
\end{tabular}
\end{center}
\end{table*}

\begin{table}
\begin{center}
\caption{Contingency table to estimate the equivalent accuracy of ECPSSR L shell
cross sections using EADL and Bearden and Burr's binding energies}
\label{tab_contl}
\begin{tabular}{|l|c|c|c|c|}
\hline
{\bf $\chi^2$ test outcome}		&{\bf ISICS 2011} &{\bf ERCS08}	\\
\hline
Pass					& 66			& 64 \\
Fail					& 18			& 20 \\
\hline
p-value Fisher test		& \multicolumn{2}{|c|}{0.854} \\
p-value Pearson $\chi^2$	& \multicolumn{2}{|c|}{0.712} \\
p-value Yates $\chi^2$	& \multicolumn{2}{|c|}{0.854} \\
\hline
{\bf $\chi^2$ test outcome}	&{\bf ISICS 2011} &{\bf LIO}	\\
\hline
Pass					& 66			& 58 \\
Fail					& 18			& 26 \\
\hline
p-value Fisher test		& \multicolumn{2}{|c|}{0.161} \\
p-value Pearson $\chi^2$	& \multicolumn{2}{|c|}{0.219} \\
p-value Yates $\chi^2$	& \multicolumn{2}{|c|}{0.160} \\
\hline
\end{tabular}
\end{center}
\end{table}

% ------------------------------------------------------------------------------
\subsection{Effect of atomic binding energies}

A detailed study \cite{tns_binding} devoted to various atomic electron binding
energy compilations in the context of Monte Carlo particle transport highlighted
the effect of these atomic parameters in the calculation of proton ionization
cross sections, regarding its accuracy with respect to experimental data.
The effects are especially visible in the resulting values of K shell cross sections.

As commented in section \ref{sec_ercs08}, ERCS08 includes an ad hoc collection
of binding energies, that does not correspond to any of the compilations used by 
major Monte Carlo systems and specialized PIXE codes.
A test was performed to estimate the impact of this collection on the accuracy
of the cross sections by running ISICS with ERCS08 binding energies: a series of
$\chi^2$ test compared the compatibility of plain ECPSSR K shell cross sections
calculated by ISICS based on either its default (Bearden and Burr
\cite{bearden}) or ERCS08 binding energies with experimental data at 0.05
significance level.
The efficiency at reproducing experimental data raised from $0.67 \pm 0.06$ with
Bearden and Burr binding energies to $0.80 \pm 0.05$ with ERCS08 binding
energies.

More extensive documentation of the sources of ERCS08 binding energies would
be beneficial to optimize the accuracy of ECPSSR cross section calculations.

% ------------------------------------------------------------------------------
\subsection{Empirical scaling}

Although successful at describing ionization cross sections over a wide energy
range relevant to PIXE experimental applications, the ECPSSR theory fails at
reproducing experimental measurements in some conditions, for instance at low
energies (below approximately 1~MeV).
The Hartree-Slater correction applied to the calculation K shell cross sections
in part overcomes these deficiencies; nevertheless, as previously pointed out,
%in \cite{tns_pixe} and in section \ref{sec_k}, 
this modeling approach exhibits other shortcomings, namely
for heavy elements at higher energies (approximately above a few MeV).

\begin{table*}
\begin{center}
\caption{Efficiency of empirically scaled and theoretical K shell ionization cross
section calculations}
\label{tab_effscaled}
\begin{tabular}{|l|ccc|cc|}
\hline
        			&\multicolumn{3}{|c|}{\bf ISICS 2011}			&\multicolumn{2}{|c|}{\bf \v{S}mit}  \\
				&ECPSSR 	& ECPSSR-HS-UA	&Scaled ECPSSR	&KIO		&KIOKC     \\
\hline
Tested elements	& 66	& 66	& 66	& 66	& 66  \\
Pass				& 44	& 51	&54 	& 47	&54 \\
Fail				& 22	& 15	&12	& 19	&12 \\
Efficiency  		& 0.67$\pm$0.06	&0.77$\pm$0.05		&0.82$\pm$0.05		&0.71$\pm$0.06 	&0.82$\pm$0.05	\\
\hline
\end{tabular}
\end{center}
\end{table*}			

An empirical correction has been developed by Paul and Muhr \cite{paul_muhr}
as a scaling function $s'_c$ to be applied to ECPSSR cross sections:
\begin{equation}
\sigma_{\text{scaled}} = s'_c\ \cdot\ \sigma_\text{ECPSSR} 
\label{eq_scaling}
\end{equation}
The parameters of the scaling function have been fitted to experimental data.

Tabulations of empirically scaled cross sections for K shell have been published
by Paul and Sacher \cite{paul_sacher}; the procedure of their calculation was
modified with respect to that adopted in \cite{paul_muhr}, but the updated
scaling function is not documented in \cite{paul_sacher}.
These tabulations are included in the PIXE data library \cite{pixe_datalib}.

Empirically scaled K shell cross sections are calculated by the KIOKC executable
in \v{S}mit's software system; this program uses the atomic binding
energies of the 1978 edition of the Table of Isotopes, consistently with those
used in \cite{paul_muhr}.

\begin{figure} [h]
\centerline{\includegraphics[angle=0,width=8.5cm]{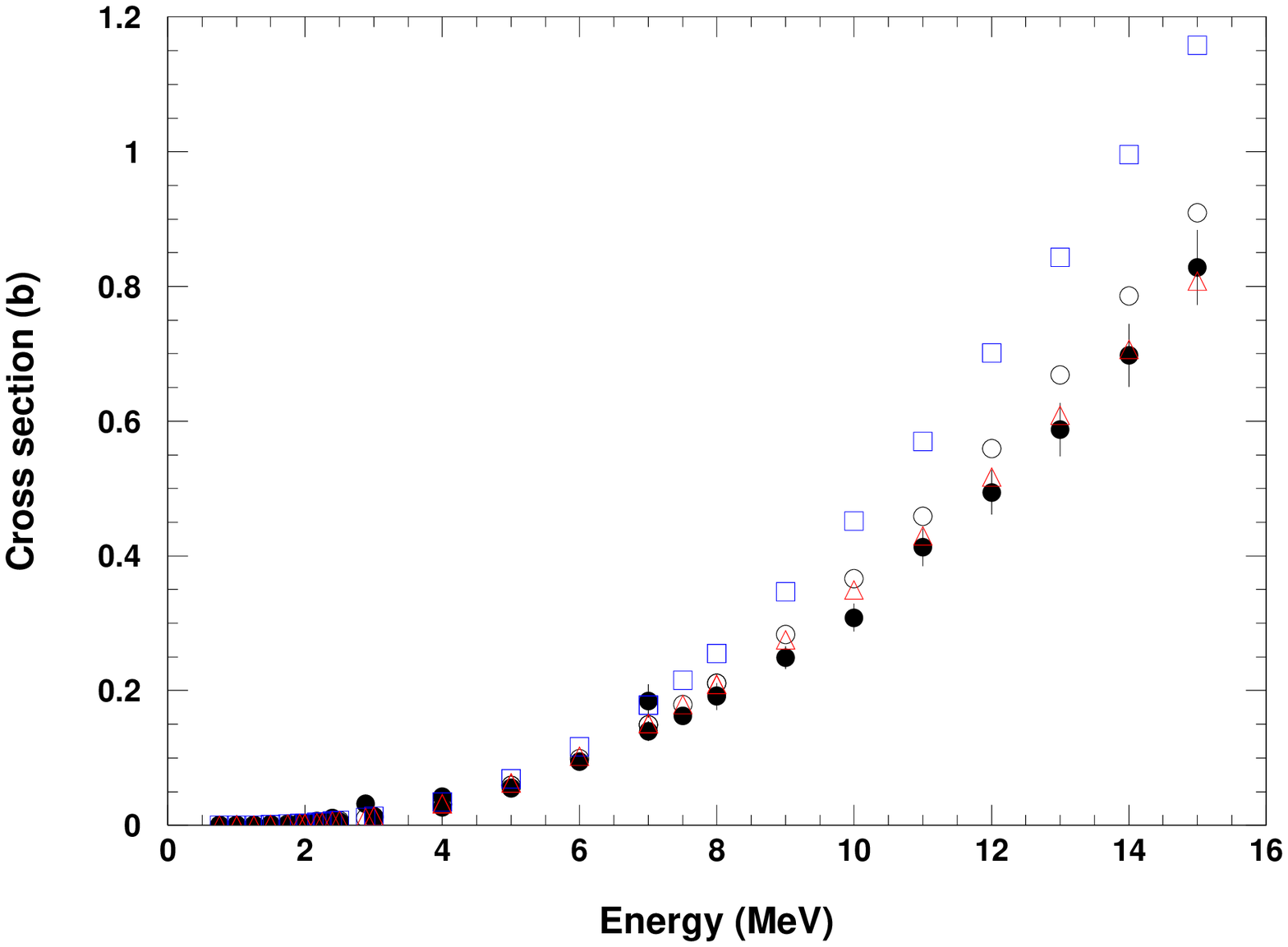}} \caption{K shell
ionization cross sections by proton impact on lead: experimental data from
\cite{paul_sacher} (black filled circles), theoretical calculations by ISICS
2011 configured with the ECPSSR-HS-UA (empty blues squares) and plain ECPSSR
(empty black circles) options, and ECPSSR values scaled by the empirical scaling
function defined in \cite{paul_muhr} (empty red triangles).}
\label{fig_scaled82}
\end{figure}

\begin{figure} [h]
\centerline{\includegraphics[angle=0,width=8.5cm]{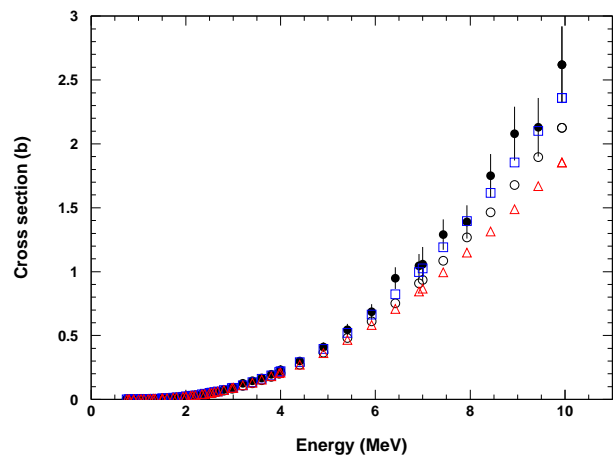}}
\caption{K shell ionization cross sections by proton impact on holmium:
experimental data from \cite{paul_sacher} (black filled circles), theoretical
calculations by ISICS 2011 configured with the ECPSSR-HS-UA (empty blue squares)
and plain ECPSSR (empty black circles) options, and ECPSSR values scaled by the
empirical scaling function defined in \cite{paul_muhr} (empty red triangles).}
\label{fig_scaled67}
\end{figure}

An implementation of Paul and Muhr's scaling function has been developed for 
use with Geant4 and is intended for release in a forthcoming Geant4 version.
Using the electron binding energies of the 1978 Table of Isotopes, it has been
verified to reproduce the values reported in  \cite{paul_muhr} for a selection of
elements and energies.
This scaling function can be applied to ECPSSR cross sections calculated by 
other generators to improve the calculation accuracy as an alternative to 
other theoretical variants, like the Hartree-Slater and United Atom corrections.

The use of this scaling function could be source of systematic effects.
The parameters defining the empirical scaling function documented in
\cite{paul_muhr} derive from a fit to the experimental data involving
calculations of ECPSSR cross sections; therefore, strictly speaking, that
empirical correction should be applied only to identically calculated ECPSSR
cross sections.
The values calculated by the three generators analyzed in this paper slightly differ 
from the ECPSSR cross sections reported in  \cite{paul_muhr} for selected 
elements and energies; however, since the discrepancies are relatively small
(e.g. less than 1\% for ECPSSR cross cross sections calculated by ISICS),
the systematic error due to the application of the empirical scaling functions to other 
ECPSSR calculations than those used in \cite{paul_muhr} would also be small,
and negligible with respect to the experimental uncertainties of K shell cross sections.

The contribution of the empirical scaling function to the cross section accuracy
has been estimated according to the procedure described in section
\ref{sec_method}; this process does not constitute a validation, since
the experimental data to which the calculated cross sections are compared are in
large part the same used for fitting the parameters of the scaling function
itself.
The usual procedure to account for fitting constraints in a $\chi^2$ test,
consisting of reducing the number of degrees of freedom accordingly, cannot be
applied to this case due to incomplete knowledge of the fit configuration.
In fact, the fits described in \cite{paul_muhr} concern groups of several
elements, while in this analysis the $\chi^2$ tests are performed for individual
elements; moreover, the experimental samples of Paul and Sacher's paper
\cite{paul_sacher}, used in this analysis, do not exactly coincide with the data
on which the scaling function drawn from Paul and Muhr's paper \cite{paul_muhr}
has been fitted, although a large fraction are common.
The ECPSSR cross sections, which are involved in the fit, are slightly different
in \cite{paul_muhr} and in this analysis.
Therefore the results reported in the following should be considered only as a
demonstration of the capability of the empirical scaling to describe the data.

The efficiency, defined as in the previous sections, 
%as the fraction of test cases for which the hypothesis of compatibility with
%experimental data is not rejected with 0.05 significance,
is listed in Table \ref{tab_effscaled} for ECPSSR cross sections calculated
by KIOKC and by ISICS scaled by Paul and Muhr's empirical function along
with the efficiency of the ISICS ECPSSR and ECPSSR-HS-UA options, and of KIO.
The gain in efficiency due to the scaling function appears substantial,
although, as illustrated in the contingency table reported in Table \ref{tab_contscaled}, the
outcome of the statistical analysis is controversial: the hypothesis of
equivalence of theoretical and empirically scaled ECPSSR cross sections is
rejected with 0.05 significance by Pearson's $\chi^2$ test, but only with 0.1
significance by Yates' $\chi^2$ test and Fisher's test.
The gain with respect to the theoretical ECPSSR-HS-UA option is not
statistically significant; nevertheless, in some experimental scenarios for
which the theory does not reproduce the measurements accurately, for instance as
illustrated in Fig. \ref{fig_scaled82}, the capability of applying empirical
scaling corrections to the theoretical cross sections could be a valuable
option.
However, within the data sample analyzed in this paper, one can also identify
cases where theoretical cross sections calculated with the Hartree-Slater and
United Atom corrections are more accurate than empirically scaled ECPSSR ones; an
example is shown in Fig.~\ref{fig_scaled67}.

\begin{table}
\begin{center}
\caption{Contingency table to estimate the equivalent accuracy of plain and
empirically scaled ISICS 2011 ECPSSR K shell cross sections}
\label{tab_contscaled}
\begin{tabular}{|l|c|c|c|c|}
\hline
{\bf $\chi^2$ test outcome}		&{\bf ECPSSR} &{\bf Scaled}	\\
\hline
Pass					& 44			& 54 \\
Fail					& 22			& 12 \\
\hline
p-value Fisher test		& \multicolumn{2}{|c|}{0.072} \\
p-value Pearson $\chi^2$	& \multicolumn{2}{|c|}{0.047} \\
p-value Yates $\chi^2$	& \multicolumn{2}{|c|}{0.073} \\
\hline
\end{tabular}
\end{center}
\end{table}

% ------------------------------------------------------------------------------
\subsection{Interpolated cross sections}

In the data driven approach to PIXE simulation described in \cite{tns_pixe}, the
cross sections calculated by analytical generators are tabulated at predefined
energies and used in Monte Carlo simulation to compute cross sections
corresponding to the energy of a given particle being tracked through matter.

A test was performed to verify if the interpolation process performed in Geant4
would significantly affect the accuracy of the theoretical cross sections.

\begin{figure} 
\centerline{\includegraphics[angle=0,width=8.5cm]{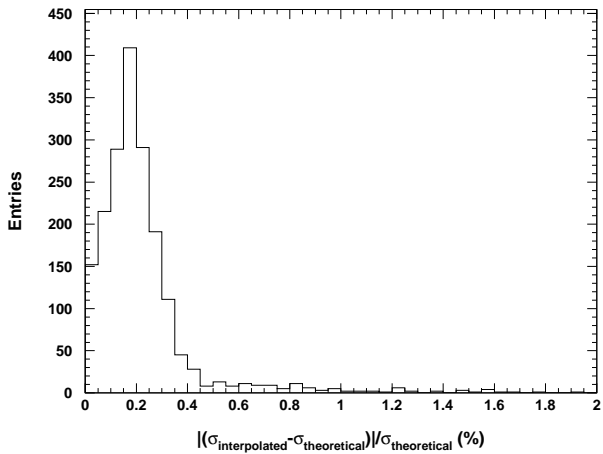}}
\caption{Absolute relative difference between interpolated and theoretical ECPSSR K shell
ionization cross sections by proton; the theoretical cross sections are produced
by ISICS, while the interpolated values are calculated by the logarithmic
interpolation algorithm encompassed in the data management component 
described in \cite{tns_pixe}, based on tabulations produced by ISICS.}
\label{fig_interpol}
\end{figure}

\begin{figure} 
\centerline{\includegraphics[angle=0,width=8.5cm]{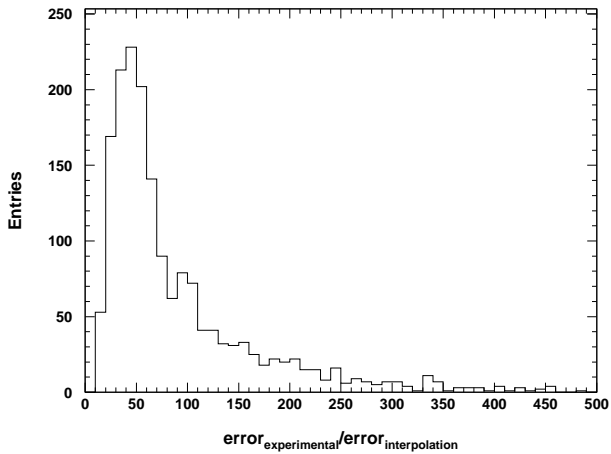}}
\caption{Ratio of K shell cross section experimental uncertainties and
interpolation errors; the interpolation error is defined as the absolute
difference between interpolated and theoretical cross sections. The theoretical
cross sections are produced by ISICS, while the interpolated values are
calculated by the logarithmic interpolation algorithm encompassed in the data
management component described in \cite{tns_pixe}, based on tabulations produced
by ISICS; the experimental uncertainties are taken from \cite{paul_sacher}.}
\label{fig_interpexp}
\end{figure}

The distribution of the relative difference between interpolated and theoretical
ECPSSR cross sections is shown in Fig. \ref{fig_interpol} for K shell cross sections;
the distributions for L subshells are similar.
The data span the energy range covered by experimental measurements; the
theoretical cross sections are generated by ISICS, while the interpolated values
are computed by logarithmic interpolation of the tabulations produced
by ISICS.
The plot includes only truly interpolated values: cross sections calculated by
the interpolation algorithm at energies corresponding to the tabulated values
are excluded, since the differences are obviously null.
The distribution can be described by a Gaussian with 0.17\% mean and
0.11\% standard deviation; the tail corresponds to regions of relatively steep variation
of the cross sections as a function of energy, mostly occurring with light targets
at low energies (below 1 MeV). 
The error due to interpolation is negligible with respect to the experimental
uncertainties of the cross sections, as can be observed in Fig.
\ref{fig_interpexp}.

Interpolated cross sections exhibit the same compatibility with experiment as
theoretically calculated ones: the outcome of the $\chi^2$ test, as rejection or
non-rejection of the null hypothesis at 0.05 significance level, is strictly
identical for ECPSSR cross sections calculated by ISICS and interpolated from
tabulations produced with ISICS, when they are compared to the same experimental
samples applying the same analysis criteria, although the test statistic assumes
slightly different numerical values for the two categories in the individual
test cases.
This result holds for cross sections interpolated from ISICS 2008 tabulations
with respect to theoretical values directly calculated by either ISICS 2008 or 2011.

From this test one evinces that the interpolation process used in Geant4
PIXE simulation \cite{tns_pixe} introduces a negligible perturbation 
to the accuracy of the cross sections calculated in the course of the simulation;
it also confirms that the updates implemented in ISICS 2011 do not modify the
accuracy of Geant4 PIXE simulation with respect to the data library distributed
with Geant4 9.4, which is based on the previous ISICS version.

%the calculated and interpolated
%cross sections, .
%
%A further analysis was performed to check the effects of the interpolation process
%regarding the compatibility with experimental data.
%For this purpose, two sets of plainwere compared to the
%experimental data samples for K and L shell: those  at
%energies corresponding to the experimental measurements, and those at predefined energies.
%The reference experimental samples and analysis criteria were the same
%for the two cross section categories.
%For all the test cases (i. e. individual experimental samples),

It is worthwhile to stress that these results concerning the interpolation of 
cross section tabulations produced by ISICS concern the data library and 
interpolation algorithm described in \cite{tns_pixe}; they would not necessarily 
hold for tabulations adopting different sampling intervals or exploiting other
interpolation algorithms.

%The comparison of the 2008 and 2010 theoretical values with experiment produces 
%identical results for the two ISICS productions: the p-values of the $\chi^2$ tests 
%have been verified to be identical up to the third decimal digit for the sets of tests.

%
%\subsection{Cross sections for Geant4-based PIXE simulation}
%
%The theoretical cross section included in data library released with Geant4 9.4
%were produced with ISICS 2008;  was investigated with respect to
%later evolutions of this generator.
%The analysis was performed in detail on K shell cross sections; 
%
%
% and whether the updated ISICS 2010 version would bring any
%significant improvements over the data library produced with the previous 2008
%version and distributed with Geant4 9.4.

% ------------------------------------------------------------------------------
%\section{PIXE Data Library}

% ------------------------------------------------------------------------------

\section{Conclusion}

Three publicly available theoretical generators of inner shell ionization by
proton impact have been analyzed to assess the state-of-the-art in the field:
the 2011 version of ISICS, the new code ERCS08 and \v{S}mit's software system
(here called KIO-LIO).

%The three generators implement calculations based on the ECPSSR theory; their
%results, although quite similar, exhibit some differences, which are due to the
%distinctive features of the three codes: calculation methods, atomic parameters
%(e.g. the electron binding energies) and software algorithms.

The accuracy achieved by the three generators has been estimated through the
statistical comparison of their results with experimental K and L shell cross
sections.
The analysis confirms quantitatively the qualitative similarity of the three
generators.
For K shell ionization, ISICS and ERCS08 exhibit the same overall efficiency at
reproducing experimental references, while the KIO system is slightly less
accurate.
For the L shell, ISICS achieves the highest efficiency and LIO the
lowest: nevertheless, the differences among the generators are not statistically
significant at 0.05 level.
One can conclude that ISICS represents the state-of-the-art among freely 
available ECPSSR generators.

Although quite similar in their overall statistical performance, the three
generators exhibit distinctive features for a few specific elements
(for instance, for silicon K shell ionization).

Their documentation in the detailed results of this paper provides guidance to
optimize the selection of cross section options in applications,
which may be sensitive to the accuracy of the results for specific targets of
experimental interest.

These results are relevant to the production of cross section tabulations
for Monte Carlo particle transport, and the selection of optimal cross section 
options in simulation applications.

The analysis also assessed the equivalent compatibility with measurements
of the 2011 and 2008 versions of ISICS; the latter was used for the production of 
the cross section data library released in Geant4 9.4 for PIXE simulation.
Moreover, it 
demonstrated that the interpolation of tabulated values does not significantly
affect the accuracy of the cross sections calculated in Geant4.

%Some features of the codes would deserve further investigation to optimize the
%accuracy of cross section data libraries: for instance, the atomic binding
%energies assembled ad hoc for ERCS08 appear to produce statistically more
%accurate cross sections than the values deriving from well established
%compilations, but the origin of these values is not fully documented, nor the
%effects that they could produce on other physical observables.
%The Hartree-Slater correction statistically improves the accuracy of K shell
%cross section calculations, but in some cases
%%for heavy elements and energies above a few MeV, 
%it produces less accurate values than plain ECPSSR
%computations: the energy range of its applicability should be optimized with the
%support of theoretical investigations.

%The analysis documented in this paper provides ground for the improvement of
%PIXE simulation with 
% data library for 

The PIXE data library has been extended to encompass tabulations produced by the
other generators analyzed in this paper in addition to those based on ISICS, so
that users can profit from the distinctive features of the three generators
identified in this study.
An updated version is intended to be released with a forthcoming version of the
Geant4 toolkit and through the RSICC distribution center.
An extension to Geant4 code is also foreseen to provide functionality for
empirically scaling ECPSSR K shell cross sections.

Apart from the calculation of cross sections for the ionization of target atoms,
which is the subject of this paper, the simulation of PIXE with Geant4 involves
the atomic relaxation, whose accuracy is documented in \cite{relax_nist}.
Other issues concerning the consistent treatment of the discrete process of
atomic relaxation along with the ionization process, which is affected by
infrared divergences, are extensively discussed in \cite{tns_pixe}.

% --------------------------------------------------------------------------

% ------------------------------------------------------------------------------
\section*{Acknowledgment}
The KIO-LIO code has been kindly provided to the authors by \v{Z}iga \v{S}mit.
The authors are grateful to Sam J. Cipolla, Vladimir Horvat and \v{Z}iga \v{S}mit for
extensive information on their codes and useful advice on how to use them,
and to Barbara Setina Bati\v{c} for valuable comments.
%The authors express their gratitude to CERN for support to the research
%described in this paper.
The CERN Library, in particular Tullio Basaglia, has provided helpful assistance
and essential reference material for this study.
%The authors thank ?? for proofreading the manuscript and valuable comments.

% ------------------------------------------------------------------------

\end{document}